# Accurate Radar Measurements of Drag Coefficients in Free Flight

Elya Courtney, Collin Morris, and Michael Courtney
Michael_Courtney@alum.mit.edu

**Abstract**
Undergraduate lab design balances several factors: 1) simple experiments connected with learning objectives 2) experiments sufficiently accurate for comparisons between theory and measurements without gaps when students ascribe discrepancies to confounding factors (imperfect simplifying assumptions, measurement uncertainties, and "human error"), and 3) experiments capturing student attention to ensure due diligence in execution and analysis.  Drag coefficient measurements are a particular challenge, though there has been some success using accurate measurements of terminal velocities.[1,2] Video shows promise in several areas of kinematics, but the number of trials in a reasonable time is limited, and analysis techniques to determine drag coefficients often include numerical integration of differential equations.[3]  Here, we demonstrate a technique with potential to measure drag coefficients to near 1% accuracy using an affordable Doppler radar system and round plastic pellets from an Airsoft launcher.

      Neglecting air resistance is common when studying projectile motion and kinematics in introductory courses.  However, careful analysis of many trajectories with sporting, recreational, and engineering interest reveals measurable discrepancies between predictions from neglecting air resistance and experimental data.  Determining drag coefficients is a necessary step to including air resistance for more accurate agreement between predictions and experiment.  A number of solvers are available on line to demonstrate the results of including air resistance if an accurate drag coefficient is available.  These solvers usually include the effects accurately by solving the applicable differential equations numerically in a manner that allows students in introductory courses to see the results without being burdened with math they may not have mastered.

      Doppler radar is an industry standard for accurate measurements of ballistic velocities and drag coefficients.  While many units are prohibitively expensive for amateur and educational uses, the LabRadar (www.mylabradar.com) device is affordable (~$600 US) and has been validated for its capabilities to accurately measure drag coefficients for numerous projectiles.[4]   However, firearms are poorly suited for classroom use due to safety considerations.  This paper reports results from measuring drag coefficients of plastic spheres 6mm in diameter and weighing much less than 1 g.  While classroom use is ultimately a matter for the instructor and school administration, Airsoft projectiles are used for game play where participants routinely shoot at each other (similar to paintball) with appropriate protective equipment.  Our experience with paintballs suggested Airsoft projectiles have better accuracy potential for precision drag measurements, being made of hard plastic; whereas, paintballs show greater variation in shape (less perfect soft spheres), size, mass, and accuracy.

Drag measurement accuracy is often dominated by shot-to-shot variations in the projectiles themselves and drag contributions from pitch and yaw of non-spherical projectiles. We designed this experiment to try two different Airsoft spheres, one with wider dimensional and mass tolerances and a clear seam in the plastic (112 mg) and one "match grade" sphere that is "double polished" to eliminate the seam and maintain better tolerances (200 mg).

LabRadar settings are configured to trigger on Doppler detection of a projectile, for projectile speeds in the archery range, and to display the Doppler-determined velocity at 3.05 m intervals from 0 to 15.24 m. A room slightly over 15 m long was used for the experiment. An outdoor setting is better for reducing radar reflections, but an indoor setting is better for eliminating wind effects.

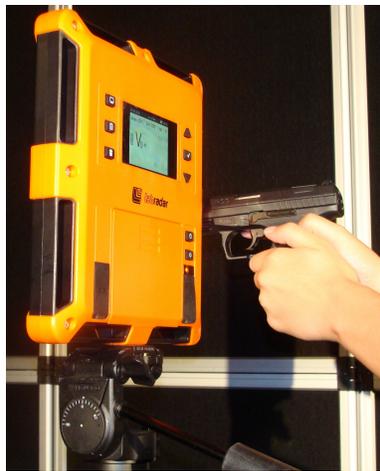

Fig 1: LabRadar shown mounted on a tripod for convenient adjustment of direction with Airsoft launcher along side.

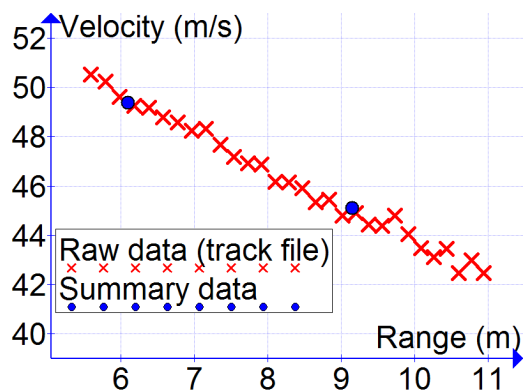

Fig 2: Velocity vs. range for one shot with the 200 mg sphere. Both raw data and summary data are shown.

Even with the launcher held next to the LabRadar as shown in Fig. 1, the sphere does not enter the radar beam and trigger the device until it is 3.05 – 6.10 m downrange, so a distance of at least 12.2 m is needed to determine near and far velocities over enough of a span to compute drag coefficients. The instrument displays extrapolated velocity measurements back to 0 m even if it does not detect the sphere for some distance downrange. One can download raw data files (Doppler velocity vs. distance and time) for each shot (USB interface) to determine the range over which Doppler velocities are directly measured (see Fig. 2) and also download a single summary file with all the shot data for a given set of conditions. This file has one row per shot, facilitating import into a spreadsheet to compute the drag coefficient for each shot using velocities over a range where they were actually measured.

The equation for computing drag coefficients from measured near and far velocities is

$$C_d = \frac{m(v_f^2 - v_i^2)}{dAv^2 \rho},$$

where $C_d$ is drag coefficient, $\rho$ is air density, $v$ is average velocity over the interval, $A$ is cross sectional area, $m$ is projectile mass, $d$ is distance between initial and final velocities, and $v_f$ and $v_i$ are final and initial velocities over the interval, respectively. This equation is functionally equivalent to equation 3 of Kagan and Nathan[5] when they use average velocity in the denominator. It is derived by setting the measured drag force over the interval (computed via the Work-Energy Theorem) equal to the theoretical drag force of an object with a given $C_d$ and solving for $C_d$.[6]

Air density can be determined from available atmospheric data with various methods.[4,5] Here, a Kestrel 4500 weather meter is used to measure ambient pressure, humidity, and temperature, and these values were entered into the JBM ballistic calculator (http://www.jbmballistics.com/ ) to determine air density with an accuracy better than 0.3%. Rather than use masses and diameters given on the product containers, masses are determined with an electronic balance, and diameters are determined with a dial caliper. The 112 mg sphere has a 1-2% variation in diameter at different measuring points and on differing spheres, so an average is used. The 200 mg sphere has no discernable variations in diameter using a dial caliper (0.025 mm resolution).

Prior experience suggested that spring launchers provide better consistency in velocities when there is not a tight fit between barrel and projectile. Though it has external similarities to a pistol, the Airsoft launcher is a plastic toy (Umarex part #2272038) with a simple spring propelling mechanism. Mean muzzle velocity with the 200 mg sphere was 57.49 m/s with a standard deviation (SD) of 0.43 m/s, and mean muzzle velocity with the 112 mg sphere was 64.74 m/s with a SD of 2.77 m/s. The 200 mg sphere had reliable Doppler velocity readings at 6.10 and 9.14 m, so velocities at these two distances were used to compute drag coefficients for each of 20 shots. The mean value of the drag coefficient was 0.396 with a standard error from the mean (SEM) of 0.9%.

The 112 mg sphere had reliable Doppler velocity readings at 3.05 and 6.10 m, so velocities at these distances were used for Cd determinations for each of 20 shots. The mean value was 0.369 with an uncertainty (SEM) of 3.2%. The greater variation in drag coefficients for this sphere are expected due to the presence of the seam, diameter variations, mass variations, and larger variations in the muzzle velocity. Teachers desiring a lab with better accuracy would do well to select Airsoft spheres that are "match grade" and "double polished."

In summary, the availability of affordable Doppler radar designed for free flight velocity measurements of projectiles makes it possible to measure drag coefficients of spheres with an accuracy of 1% or so if some care is taken to select uniform spheres. These measured drag coefficients are consistent with expectations for these Mach ranges and Reynolds numbers, as is the observation that the less uniform sphere has less drag than the more perfect sphere.

**About the Authors**
Elya attends the University of Georgia majoring in Chemistry on a full academic scholarship.

Collin Morris is a Mechanical Engineering major at Georgia Tech with a particular interest in aerodynamics.

Michael Courtney earned a PhD in Physics from MIT in 1995 and has served in a number of faculty positions, including at the US Air Force Academy. He founded BTG Research in 2001 to study ballistics with applications in military and law enforcement, but he also delights in undergraduate applications of ballistics in research and physics labs.